\documentstyle[11pt,newpasp,twoside,epsf,amssymb]{article}
\renewcommand{\vec}[1]{\mbox{\boldmath $\displaystyle #1$}}
\newcommand{\grad}{\vec{\nabla}}

\newcommand{\vcross}{\vec{\times}}

\newcommand{\curl}{\grad\vcross\,}
\newcommand{\be}{\begin{eqnarray}}
\newcommand{\ee}{\end{eqnarray}}

\newcommand{\nat}{Nature}
\newcommand{\pre}{Phys.~Rev.~E}

\begin{document}
\title{Magnetic Field Evolution During Neutron Star Recycling}

\author{Andrew Cumming}
\affil{Department of Astronomy and Astrophysics, University of California, Santa Cruz, CA 95064}

\begin{abstract}
I describe work on two aspects of magnetic field evolution relevant for the ``recycling'' scenario for making millisecond radio pulsars. First, many of the theoretical ideas for bringing about accretion-induced field decay rely on dissipation of currents in the neutron star crust. I discuss field evolution in the crust due to the Hall effect, and outline when it dominates Ohmic decay. This emphasises the importance of understanding the impurity level in the crust. Second, I briefly discuss the progress that has been made in understanding the magnetic fields of neutron stars currently accreting matter in low mass X-ray binaries. In particular, thermonuclear X-ray bursts offer a promising probe of the magnetic field of these neutron stars.
\end{abstract}

\section{The Crust as a Site for Dissipation of Currents}

While spin up is naturally expected from accretion of angular momentum, magnetic field decay in an accreting neutron star is much less well-understood (see Bhattacharya \& Srinivasan 1995 for a review). The timescales for evolution of currents in the core of a neutron star is extremely long (Baym, Pethick, \& Pines 1969; Goldreich \& Reisenegger 1992, hereafter GR92). However, short decay times ($\lesssim 10^7$ years) are possible in the neutron star crust (as emphasised very early on, e.g.~Ewart et al.~1975), and this underlies one class of models for accretion-induced field decay.

One idea is to place the currents supporting the stellar magnetic field entirely in the crust. This might be the case if the field is generated from thermomagnetic effects, for example (Blandford, Applegate, \& Hernquist 1983). The evolution due to Ohmic decay is then straightforward to calculate, and has been followed in many papers (Urpin, Geppert, \& Konenkov 1997, Konar \& Bhattacharya 1997, and references therein). The role of accretion is to heat the crust, reducing the electrical conductivity. The amount by which the field decays depends on the accretion lifetime, depth of the currents, and how much current is advected and subsequently ``frozen'' into the superconducting core. Thermomagnetic processes may also destroy field in the reheated crust (Blondin \& Freese 1986). 

A second proposal is that as a radio pulsar spins down, outward moving vortices in the superfluid and superconducting core push magnetic fluxoids into the crust, where magnetic energy dissipates (Srinivasan et al.~1990; Konar \& Bhattacharya 1999). This model predicts changes in alignment of the magnetic field and spin axis (perhaps involving ``plate tectonics'' of the crust, e.g.~Chen \& Ruderman 1993). 

The importance of understanding evolution of currents in the crust for these different models motivated our recent study (Cumming, Arras, \& Zweibel 2004, hereafter CAZ04). Whereas calculation of Ohmic decay is straightforward, the evolution of currents in the crust is complicated by the non-linear Hall effect, first studied by Jones (1988) and GR92. This complex process in fact has very simple underlying physics: in the crust, where the ions are held fixed in the solid lattice, {\em the magnetic field is frozen into the electron fluid}. To see this, combine the Hall electric field (familiar from the simple laboratory experiment) $\vec{E}_H=\vec{v}_e\vcross\vec{B}/c$, where $\vec{v_e=J/n_ee}$ is the electron velocity, $\vec{J}$ is the current density, and $n_e$ is the electron density, with Faraday's law, giving $\partial\vec{B}/\partial t=-c\curl{\vec{E}_H}=-\curl{\vec{v_e}\vcross\vec{B}}$. Therefore the field $\vec{B}$ evolves due to the currents $\vec{J}\propto\curl\vec{B}$. An initial dipole field will spontaneously ``twist'' and ``buckle'' as the currents distort the dipole field lines inside the star, generating higher order multipoles (see CAZ04 \S 4.1).

How fast is this evolution, and when is the Hall effect important? The Hall effect always dominates for magnetar-strength fields $\sim 10^{14}$--$10^{15}\ \mathrm{G}$. However, for $B<10^{13}\ \mathrm{G}$, the importance of the Hall effect depends sensitively on the composition of the crust. A simple timescale estimate is given by writing the Hall time across a pressure scale height at the base of the crust, $t_{\mathrm{Hall}}\sim 10\ \mathrm{Myrs}/B_{12}$ where $B_{12}$ is $B$ in units of $10^{12}\ \mathrm{G}$ (see also GR92). The Ohmic decay time is $t_{\mathrm{Ohm}}\approx 2\ \mathrm{Myrs}/T_8^2$ when phonons dominate the electrical conductivity (temperatures $T_8=T/10^8\ \mathrm{K}\gtrsim 1$), and $t_{\mathrm{Ohm}}\approx 6\ \mathrm{Myrs}/Q$ when impurities dominate ($T_8\lesssim 1$), where the impurity parameter $Q$ measures the level of impurities in the crystal lattice (e.g.~Itoh \& Kohyama 1993). Figure 1 shows curves of $t_{\rm Ohm}=t_{\rm Hall}$ for different $Q$, and rough locations of different types of neutron star\footnote{We choose a density $\rho=10^{14}\ \mathrm{g\ cm^{-3}}$ (near the base of the crust) in Figure 1. The ratio $t_{\rm Ohm}/t_{\rm Hall}\propto \rho^{1/6}$ for phonon scattering, and $\propto 1/\rho^{2/3}$ for impurity scattering.}. 

Figure 1 shows that recent work on the crust composition in both isolated and accreting neutron stars has a direct impact on our understanding of field evolution in the crust. Accreting neutron stars are hot ($T_8\gtrsim 1$ for accretion rates $\dot M\gtrsim 10^{-11}\ M\odot\ \mathrm{yr^{-1}}$), but also likely have impure crusts $Q\gtrsim 1$, since their crusts are replaced by a mixture of heavy elements made by hydrogen and helium burning on the surface of the neutron star (Schatz et al.~1999 found $Q\sim 100$). Therefore, the Hall effect is never important for an accreting star, and the Ohmic time remains short even after accretion switches off. The original estimate of $Q$ for an isolated neutron star by Flowers \& Ruderman (1977) gave $Q\sim 10^{-3}$, in which case Hall effects dominate in a radio pulsar which has cooled to $T_8\lesssim 1$, and a cascade to small lengthscales is expected. However, recent calculations by Jones (2004, and references therein) give $Q\gtrsim 1$ for an isolated neutron star, in which case an extensive Hall cascade is less likely, with Ohmic decay dominating for $B_{12}\lesssim Q\rho_{14}^{2/3}$.

When Hall effects dominate the evolution, the field is expected to grow more and more complex. GR92 suggested that a turbulent ``Hall cascade'' would transport magnetic energy to small scales where Ohmic dissipation is very efficient. For a constant electron density, this cascade has now been confirmed in numerical simulations (Biskamp et al.~1999). However, unfortunately it is difficult to draw firm conclusions about this evolution. Many fascinating physics issues remain to be resolved, including the nature of the cascade (perhaps strong and anisotropic), the behavior of Hall waves in the presence of a steep density gradient and at the boundaries of the crust, and the response of the crust to the wave for strong fields (Rheinhardt \& Geppert 2000; Vainshtein, Chitre, \& Olinto 2000; CAZ04). These issues are particularly important for magnetars, since the time for the Hall effect to operate is very short (and always dominates Ohmic decay, independent of $Q$), and perhaps drives the magnetic field decay believed to power these sources (Thompson \& Duncan 1996; see Arras, Cumming, \& Thompson 2004 for a recent model).

\begin{figure}
\begin{center}
\epsfxsize=3.5in
\epsfbox{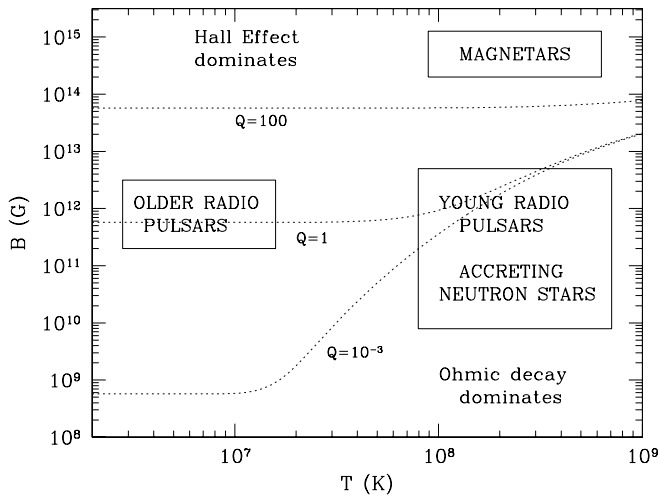}
\end{center}
\caption{Regimes of dissipation of crustal currents: Ohmic decay vs.~Hall effect for different neutron stars (adapted from CAZ04). The dotted lines show the magnetic field $B$ above which $t_{\rm Hall}<t_{\rm Ohm}$ as a function of temperature in the crust (for a density $\rho=10^{14}\ {\rm g\ cm^{-3}}$), and different impurity parameters $Q$. Boxes mark the approximate location of different types of neutron star. }
\end{figure}

%==========================================
\section{The Magnetic Fields of LMXB Neutron Stars}

Discoveries with the Rossi X-ray Timing Explorer (RXTE) have led to spin measurements for several neutron stars in low mass X-ray binaries (LMXBs), providing important confirmation that the neutron stars in these systems are spinning rapidly (spin frequencies $185$--$619\ \mathrm{Hz}$, Chakrabarty et al.~2003). Five systems show persistent X-ray pulsations (Wijnands 2004) indicating magnetically-channelled accretion with a stellar magnetic field of $B\sim 10^8$--$10^9\ \mathrm{G}$ (Psaltis \& Chakrabarty 1999), precisely as expected for millisecond pulsar (MSP) progenitors. Oscillations during Type I X-ray bursts give spin measurements for a further 11 systems (Muno 2004). These bursts are due to unstable thermonuclear burning of accreted hydrogen and helium on the neutron star surface, and the oscillations are thought to arise because the burning is not spherically symmetric, leading to modulation of the X-rays at the rotation period of the star.

Burst oscillations are a potential probe of the neutron star magnetic field. The burst oscillation frequency drifts slightly by a few Hz during the $\sim 10\ \mathrm{s}$ burst. This is interpreted as some kind of fluid motion over the neutron star surface, either a global drifting due to angular momentum conservation as the layer heats up and expands outwards, drift of a rotationally-supported hotspot, or perhaps oscillation modes excited in the neutron star ocean (see Muno 2004 for a review and references). Cumming \& Bildsten (2000) (CB00) pointed out that even a weak magnetic field could interfere with these motions on the burst timescale, magnetic tension acting to brake the flow. 

An estimate for the magnetic braking timescale is the Alfven crossing time, which is $\approx 0.01\ {\rm s}\ (B/10^8\ {\rm G})^{-1}$, much shorter than the burst duration (CB00). If this simple argument is correct, the observed drift over several seconds implies a much weaker magnetic field in the burning layers than expected for these presumed MSP progenitors. Conversely, a simple prediction for the persistently pulsating sources (which show direct evidence of such fields) would be that burst oscillations from these sources should not show large drifts (Cumming, Zweibel, \& Bildsten 2001) (CZB01). In fact, the persistent pulsator SAX J1808.4-3658 shows a large frequency drift during the rise of a Type I burst, but $\sim 10$ times faster than other burst oscillation sources. Chakrabarty et al.~(2003) proposed that this fast drifting is an indication of a stronger magnetic field in this system than most burst oscillation sources. 

An important question to answer is what makes the 5 accreting pulsars different from most LMXBs which have shown no evidence for persistent pulsations despite extensive searches.   One possibility is that the magnetic field is the expected $10^8$--$10^9$G in most sources, but is also ``screened'' by accretion. CZB01 showed that accretion at rates $\dot M\gtrsim 10^{-10}\ M_\odot\ {\rm yr^{-1}}$ is rapid enough to prevent Ohmic diffusion magnetizing the freshly accreted material. Instead, the field is pushed into the neutron star ocean, where screening currents develop and act to reduce the field in the outermost layers. This fits nicely with the observation that the five accreting pulsars are in weak transient systems, with unusually low time-averaged accretion rates ($\langle\dot M\rangle\sim 10^{-11}\ M_\odot\ {\rm yr^{-1}}$). Accretion at such low rates gives plenty of time the magnetic field to emerge by Ohmic diffusion, making screening ineffective. A screened interior field would also explain why most burst oscillations show slow frequency drifts. However, much more work is needed on the physics in the burning layer. For example, magnetic fields are likely ubiquitous: small scale field is expected to be produced by convective motions (Spitkovsky et al.~2002), and thermomagnetic drift can act to transport the underlying stellar field upwards (Cumming \& Zweibel 2003).

\acknowledgements 
I thank Phil Arras for comments on the manuscript. This work was supported by NASA Hubble Fellowship grant HF-01138 awarded by the Space Telescope Science Institute, operated for NASA by the Association of Universities for Research in Astronomy, Inc., under contract NAS 5-26555.


\begin{references}

\noindent
Arras, P., Cumming, A., \& Thompson, C. 2004, ApJ in press (astro-ph/0401561)

\noindent
Baym, G., Pethick, C., \& Pines, D.1969, \nat, 224, 674

\noindent
Bhattacharya D., \& Srinivasan, G. 1995, in X-Ray Binaries,
ed. W. H. G. Lewin, J. van Paradijs, \& E. P. J. van den Heuvel
(Cambridge: Cambridge University Press), 495

\noindent
Biskamp, D., Schwarz, E., Zeiler, A., Celani, A., \& Drake,
J. F. 1999, Phys. Plasmas, 6, 751

\noindent
Blandford, R.~D., Applegate, J.~H., \& Hernquist, L.\ 1983, \mnras, 204, 1025 

\noindent
Blondin, J.~M.~\& Freese, K.\ 1986, \nat, 323, 786 

\noindent
Chakrabarty, D., Morgan, E.~H., Muno, M.~P., Galloway, D.~K., Wijnands, R., van der Klis, 
M., \& Markwardt, C.~B.\ 2003, \nat, 424, 42 

\noindent
Chen, K.~\& Ruderman, M.\ 1993, \apj, 408, 179 

\noindent
Cumming, A., Arras, P., \& Zweibel, E. G. 2004, \apj, in press (astro-ph/0402392) (CAZ04)

\noindent
Cumming, A., \& Bildsten, L. 2000, \apj, 544, 453 (CB00)

\noindent
Cumming A., \& Zweibel E.~G. 2003, AAS HEAD meeting abstract 42.01

\noindent
Cumming A., Zweibel E.~G., \& Bildsten L. 2001, \apj, 557, 958 (CZB01)

\noindent
Ewart, G.~M., Guyer, R.~A., \& Greenstein, G. 1975, \apj, 202, 238

\noindent
Flowers, E., \& Ruderman, M. A. 1977, \apj, 215, 302

\noindent
Goldreich P., \& Reisenegger A. 1992, \apj, 395, 250 (GR92)

\noindent
Itoh, N.~\& Kohyama, Y.\ 1993, \apj, 404, 268 

\noindent
Jones, P.~B.\ 1988, \mnras, 233, 875 

\noindent
Jones, P.~B.\ 2004, \mnras\ in press (astro-ph/0403400)

\noindent
Konar, S.~\& Bhattacharya, D.\ 1997, \mnras, 284, 311 

\noindent
Konar, S.~\& Bhattacharya, D.\ 1999, \mnras, 308, 795 

\noindent
Muno, M.~P. 2004, in Proc. of "X-ray Timing 2003: Rossi and  Beyond", eds. P. Kaaret, F. K. Lamb, \& J. H. Swank (Melvill, NY; AIP)

\noindent
Psaltis, D.~\& Chakrabarty, D.\ 1999, \apj, 521, 332 

\noindent
Rheinhardt, M., \& Geppert, U. 2000, \prl, 88, 101103

%\noindent
%Romani, R. W. 1990, \nat, 347, 741

\noindent
Srinivasan, G., Bhattacharya, D., Muslimov, A.~G., \& Tsygan, A.~J.\ 1990, Current Science, 59, 31

\noindent
Thompson, C.~\& Duncan, R.~C.\ 1996, \apj, 473, 322 

\noindent
Urpin, V., Geppert, U., \& Konenkov, D.\ 1998, \mnras, 295, 907 

\noindent
Vainshtein, S. I., Chitre, S. M., \& Olinto, A. V. 2000, \pre, 61, 4422

\noindent
Wijnands, R. 2004, in Proc. of "X-ray Timing 2003: Rossi and  Beyond", eds. P. Kaaret, F. K. Lamb, \& J. H. Swank (Melvill, NY; AIP)

\noindent
Young, E. J., \& Chanmugam, G. 1995, \apj, 442, L53

\end{references}
\end{document}